\documentclass[article]{elsarticle}
\usepackage[T1]{fontenc}
\usepackage{caption}
\usepackage{subcaption}
\usepackage{lineno,hyperref}
\usepackage{amsmath}
\usepackage{amssymb}
\usepackage{mathtools}
\usepackage{tablefootnote}
\usepackage{soul}
\usepackage{ulem}
\usepackage{color}
\usepackage{float}
\usepackage{makecell}
\usepackage[usenames,dvipsnames,svgnames,table]{xcolor}
\usepackage{indentfirst}
\usepackage{leftidx}
\usepackage{slashed}

\begin{document}
	\begin{frontmatter}
		%\title{On Neutrino Oscillations and the Hubble Tension}
		\title{Using Neutrino Oscillations to Measure $H_0$}
		\author[ICC,UB]{Ali Rida Khalifeh\corref{cor1}}
		\ead{ark93@icc.ub.edu}
		\cortext[cor1]{Corresponding author}
		
		\author[ICC,ICREA]{Raul Jimenez}
		\ead{raul.jimenez@icc.ub.edu}

		\address[ICC]{ICC, University of Barcelona, Marti  i Franques, 1, E-08028 Barcelona, Spain.}
		\address[UB]{Dept. de  Fisica Cuantica y Astrofisica, University of Barcelona, Marti  i Franques 1, E-08028 Barcelona, Spain.}
		\address[ICREA]{ICREA, Pg. Lluis Companys 23, Barcelona, E-08010, Spain.}

		\begin{abstract}
			The tension between late and early universe probes of today's expansion rate, the Hubble parameter $H_0$, remains a challenge for the standard model of cosmology $\Lambda$CDM. There are many theoretical proposals  
			%have been made so far
			to remove the tension, with work still needed on that front. However, by looking at new probes of the $H_0$ parameter one can get new insights that might ease the tension. Here, we argue that neutrino oscillations could be such a probe. We expand on previous work and study the full three-flavor neutrino oscillations within the $\Lambda$CDM paradigm. We show how the oscillation probabilities evolve differently with redshift for different values of $H_0$ and neutrino mass hierarchies. We also point out how this affects neutrino fluxes which, from their measurements at neutrino telescopes, would determine which value of $H_0$ is probed by this technique, thus establishing the aforementioned aim.
		\end{abstract}
	\end{frontmatter}
	\section{Introduction}
	
	The Hubble tension, the discrepancy between early and late universe measurements of the Hubble parameter $H_0$, is still persisting~\cite{Riess:Htension,VerdeTreuRiess}. Early universe probes are mainly from Cosmic Microwave Background (CMB) experiments, such as the ones from~\cite{Planck2018,DES:Y3Res}. This parameter is determined assuming $\Lambda$ Cold Dark Matter ($\Lambda$CDM) as the fiducial cosmological model, combined with measurements independent from it. On the other hand, late universe ones use the local distance ladder method~\cite{Jackson:Distance-Ladder,Rowan-Robinson:DistanceLadder} on Cepheids, type-Ia supernovae and tip of the red giant branch in a way independent from the cosmological model~\cite{Riess:accelerating1,H0LiCOW:HTension}.
	
	To solve this tension, several theoretical models have been proposed, including early dark energy~\cite{EDE1,EDE2} and modified gravity~\cite{MG-HTension} (see~\cite{Olympiad}  for a recent and thorough review on the subject). However, one can gain new insight on this tension by developing new observables, being late or early universe ones, that are affected by today's expansion rate.
	
	In this work, we build on previous ones~\cite{Khalifeh:2020bdg,Khalifeh:2021ree} and show the possibility of using neutrino oscillations as a new probe for the Hubble tension. Although the latter has been looked at in connection with neutrinos previously~\cite{Neutrino-Hubble}, our approach is quite different. We consider a system of three-flavor neutrinos, $(\nu_{e},\nu_{\mu},\nu_{\tau})$, traveling in a flat Friedmann-Lema\^itre-Robertson-Walker (FRW) spacetime with a cosmological constant Dark energy (DE) $\Lambda$. By studying the transition probabilities' evolution from one flavor to another as a function of redshift, we show how different values of $H_0$ affect the detected neutrino fluxes, making the latter a potential probe for $H_0$. In our analysis, we consider different initial conditions (ICs) for neutrino flavor decomposition, and distinguish between their mass hierarchies.
	
	It should be noted that there have been a great deal of work in the literature done on neutrinos as spinors in curved spacetime. We direct the interested reader to a few of them and references therein~\cite{Blasone:2018ktu,Time-EnergyUncertainty,Buoninfante:2019der,Capolupo,Cardall-Fuller,Kaplan-Nu,Kamionkowski_Nu_Lorentz,Nu_DE_Data,Mohseni,MohseniSadjadi:2017jne,Luciano:2021gdp}.
	
	The organization of the paper is as follows: we briefly present the necessary principles and equations for the analysis in section~\ref{Sec:Equations}. Then, we present and discuss the main results, given as triangular plots, what is called ternary diagrams, and fluxes' evolution with redshift, in section~\ref{Sec:Results}. We finish with some concluding remarks in section~\ref{Sec:Conc}. 
	
	We use units in which $\hbar=c=1$ and a metric signature $(-,+,+,+)$. Moreover, data from~\cite{Planck2018} is used to get an early universe (EU) value of $H_0$, what we call $H_0^{EU}=2.13\times10^{-33}h^{\text{EU}}$eV, where $h^{\text{EU}}=0.674$. In addition to that, matter and DE density parameters $\Omega_{m(\Lambda)}=8\pi G/(3H_0^2)\rho_{m(\Lambda)}$, where $\rho_m$, $\rho_{\Lambda}$ are the energy densities of matter and DE, respectively, are also taken from~\cite{Planck2018}. For the late universe (LU) value of $H_0$, $H_0^{LU}$, we use results from~\cite{Riess:2019cxk}, which gives $H_0^{LU}=2.13\times10^{-33}h^{\text{LU}}$ eV, with $h^{\text{LU}}=0.740$. 
	
	Notation wise, neutrino flavor states will be denoted by Greek indices, while Latin ones denote mass eigenstates.
	\section{Neutrinos in flat FLRW Universe}
	\label{Sec:Equations}
	
	In this section, we will follow a practical approach in which we briefly describe the relevant equations and principles needed for the case under study. We refer the unfamiliar reader to~\cite{Khalifeh:2020bdg,Khalifeh:2021ree} and references therein for a more thorough derivation.
	
	In the concordance $\Lambda$CDM model, spacetime is best described by a flat FLRW metric $g_{\mu\nu}$, given by the line element
	\begin{equation}
	ds^2=g_{\mu\nu}dx^{\mu}dx^{\nu}=-dt^2+a^2(t)\big(dr^2+r^2d\theta^2+r^2\sin^2\theta d\phi^2\big)
	\label{Eq:Metric}
	\end{equation}
	in terms of cosmic time $t$ and spherical coordinates $\{r,\theta,\phi\}$. Moreover, $a(t)$, the scale factor, is independent of spatial coordinates due to homogeneity and isotropy of FRW. Following the usual machinery in Cosmology~\cite{Dodelson,Piattella:2018hvi}, one gets the first Friedmann equation
	\begin{equation}
	H^2(z)=\frac{8\pi G}{3}\big(\rho_m+\rho_{\Lambda}\big)=H_0^2\bigg(\Omega_m(1+z)^3+\Omega_{\Lambda}\bigg)
	\label{Eq:Friedmann}
	\end{equation} 
	where $1+z=a_0/a$ is the redshift, with $a_0$ being today's value of $a(t)$.
	
	This is what we will be needing from the gravity side. On the neutrino part, to study their oscillations in curved spacetime, we are mainly interested in the transition amplitude between two flavor states $|\nu_{\alpha}\rangle$ and $|\nu_{\beta}\rangle$, $\Psi_{\alpha\beta}$, from which we get the oscillation probability $P_{\alpha\beta}$:
	\begin{equation}
	\Psi_{\alpha\beta}\equiv\langle\nu_{\beta}|\nu_{\alpha}\rangle\Rightarrow P_{\alpha\beta}=|\Psi_{\alpha\beta}|^2.
	\label{Eq:Trans_ampl_Prob}
	\end{equation}
	As was shown previously~\cite{Khalifeh:2021ree}, $\Psi_{\alpha\beta}$ evolves with the affine parameter $\lambda$ for the case of $\Lambda$CDM as:
	\begin{equation}
	i\frac{d}{d\lambda}\Psi_{\alpha\beta}=\frac{1}{2}\mathcal{M}_f^2\Psi_{\alpha\beta},
	\label{Evol_Trans_Amp}
	\end{equation}
	where 
	\begin{align}
	\mathcal{M}_f^2 &= U\begin{pmatrix}
	m_1^2 & 0 & 0\\
	0 & m_2^2 &0\\
	0 & 0 & m_3^2
	\end{pmatrix}U^{\dagger}
	\label{Eq: Mass_Matrix}
	\end{align} 
	is the square of the vacuum mass matrix in flavor space. In the above equation, $m_i$, for $i=1,2,3$, are the eigenvalues of the neutrino mass states, $|\nu_i\rangle$, and $U_{\alpha j}$ is the  Pontecorvo-Maki-Nakagawa-Sakata(PMNS) matrix for neutrino mixing~\cite{Pontecorvo:1967fh,Maki-Nakagawa-Masami}. More explicitly, the latter can be written in terms of mixing angles $\theta_{ij}$, for $\{i,j\}=1,2,3$, as~\cite{Review-Particle-Physics}
	\begin{align}
	U &=\begin{pmatrix}
	c_{12}c_{13} & s_{12}c_{13} & s_{13}e^{-i\delta}\\
	-s_{12}c_{23}-s_{13}s_{23}c_{12}e^{i\delta} & c_{12}c_{23}-s_{12}s_{23}s_{13}e^{i\delta} & s_{23}c_{13} \\
	s_{12}s_{23}-s_{13}c_{12}c_{23}e^{i\delta} & -s_{23}c_{12}-s_{12}s_{13}c_{23}e^{i\delta} & c_{13}c_{23}
	\end{pmatrix}
	\end{align}
	where $c_{ij}\equiv\cos\theta_{ij}$, $s_{ij}\equiv\sin\theta_{ij}$ and $\delta$ is the Charge Conjugation-Parity (CP) violating phase. Incidentally, here we are considering neutrinos to be of the Dirac type, hence there is one CP violating phase (see~\cite{CP-Violation,DiracVsMajorana} for a review on CP violation and the nature of neutrinos).
	
	If we start with an initial state $|\nu_{\alpha}\rangle$, then eq.~\eqref{Evol_Trans_Amp} can be written explicitly as
	\begin{align}
	i\frac{d}{d\lambda}\begin{pmatrix}
	\Psi_{\alpha e} \\
	\Psi_{\alpha\mu}\\
	\Psi_{\alpha\tau}
	\end{pmatrix} &=\frac{1}{2}U\begin{pmatrix}
	m_1^2 & 0 & 0\\
	0 & m_2^2 &0\\
	0 & 0 & m_3^2
	\end{pmatrix}U^{\dagger}
	\begin{pmatrix}
	\Psi_{\alpha e} \\
	\Psi_{\alpha\mu}\\
	\Psi_{\alpha\tau}
	\end{pmatrix}
	\label{Evol_Eq_Expl}
	\end{align}
	for the transition of $\alpha$ to any of the three flavors $e, \mu$ and $\tau$. By defining
	\begin{equation}
	\Phi_{\alpha}\equiv U^{\dagger}\Psi_{\alpha},
	\label{Def_Phi}
	\end{equation}
	eq.~\eqref{Evol_Eq_Expl} becomes, after multiplying it with $U^{\dagger}$ from the left,
	\begin{align}
	i\frac{d}{d\lambda}\begin{pmatrix}
	\Phi_{\alpha 1} \\
	\Phi_{\alpha 2}\\
	\Phi_{\alpha 3}
	\end{pmatrix} &=\frac{1}{2}\begin{pmatrix}
	m_1^2 & 0 & 0\\
	0 & m_2^2 &0\\
	0 & 0 & m_3^2
	\end{pmatrix}
	\begin{pmatrix}
	\Phi_{\alpha 1} \\
	\Phi_{\alpha 2}\\
	\Phi_{\alpha 3}
	\end{pmatrix}.
	\label{Evol_Phi}
	\end{align}
	Here, we have used the unitarity condition of the PMNS, $U^{\dagger}U=I$, where $I$ is the $3\times3$ identity matrix. As one can see, by going from eq.~\eqref{Evol_Eq_Expl} to eq.~\eqref{Evol_Phi} we have simply changed from flavor to mass basis. This will make it easier to solve the evolution equation, and then we can simply transform back to the flavor basis by the inverse of~\eqref{Def_Phi}.
	
	The solution to each $\Phi_{\alpha i}$ is
	\begin{equation}
	\Phi_{\alpha i}(\lambda)=\Phi^{\text{ini}}_{\alpha}e^{-i\frac{1}{2}m_i^2\Delta\lambda}
	\label{Sol_Phi}
	\end{equation}
	where $\Phi^{\text{ini}}_{\alpha}$ is the initial neutrino composition at emission in mass basis, and~\cite{Khalifeh:2021ree}
	\begin{equation}
	\Delta\lambda\equiv\frac{1}{E_0}\int_{0}^{z_e}\frac{dz}{H(z)(1+z)^2},
	\end{equation}
	with $E_0$ being the detected neutrino energy on Earth (corresponding to $z=0$), $z_e$ is the source's redshift and $H(z)$ is given in eq.~\eqref{Eq:Friedmann}. 
	
	Finally, to get the probability $P_{\alpha\beta}$, there are three steps that need to be done. First, starting from an initial neutrino flavor composition, $\Psi^{\text{ini}}$, we apply eq~\eqref{Def_Phi} to get $\Phi^{\text{ini}}_{\alpha}$. Second, we plug the latter in the  solution eq.~\eqref{Sol_Phi} and apply the inverse of eq.~\eqref{Def_Phi} to get the evolution of $\Psi_{\alpha\beta}$. Third, we take the modulus square of that to get an expression for the $\alpha\rightarrow\beta$ transition probability of the form:
	\begin{equation}
	P_{\alpha\beta}=\delta_{\alpha\beta}+\sum_{i<j}^{}\bigg[a_{\alpha\beta;ij}\sin^2\bigg(\frac{\Delta m_{ij}^2\Delta\lambda}{4}\bigg)+b_{\alpha\beta;ij}\sin\bigg(\frac{\Delta m_{ij}^2\Delta\lambda}{2}\bigg)\bigg]
	\label{Probability}
	\end{equation}
	%\begin{align}
	%P_{\alpha\beta} &= \delta_{\alpha\beta}+a^1_{\alpha\beta}\sin^2\bigg(\frac{\Delta m_{21}^2\Delta\lambda}{4}\bigg)+a^2_{\alpha\beta}\sin^2\bigg(\frac{\Delta m_{31}^2\Delta\lambda}{4}\bigg)+a^3_{\alpha\beta}\sin^2\bigg(\frac{\Delta m_{32}^2\Delta\lambda}{4}\bigg)\nonumber \\
	%& +b^1_{\alpha\beta}\sin\bigg(\frac{\Delta m_{21}^2\Delta\lambda}{2}\bigg)+b^2_{\alpha\beta}\sin\bigg(\frac{\Delta m_{31}^2\Delta\lambda}{2}\bigg)+b^3_{\alpha\beta}\sin\bigg(\frac{\Delta m_{32}^2\Delta\lambda}{2}\bigg),
	%\label{Probability}
	%\end{align}
	where $\delta_{\alpha\beta}$ is the Kronecker delta, $\Delta m_{ij}^2\equiv m_i^2-m_j^2$, and the $a_{\alpha\beta;ij}$s\footnote{To avoid confusion, we note that the semicolon does not correspond to any kind of derivative, but it's used just to seperate flavor from mass indicies.} and $b_{\alpha\beta;ij}$s are numerical factors resulting from different combinations of PMNS components. In particular, this combination depends on which states $\alpha$ and $\beta$ are being considered (see eq. (13.9) in~\cite{Giunti:2007ry} for the equivalent form in Minkowski spacetime).
	
	From here, we can apply the above machinery to several initial conditions and see how it affects the probability's evolution with redshift, in addition to that of the flux, which will be the subject of the next section.
	
	\section{Observational Results for Different Initial Conditions}  
	\label{Sec:Results}
	
	The space of initial conditions (ICs) for neutrino oscillations, i.e. initial decomposition,  has many elements. However, there are three that are more relevant observationally: Neutron decay (ND), Muon damping (MD) and Pion decay (PD). In the representation $(\nu_e:\nu_{\mu}:\nu_{\tau})$ for the initial ratios, these three conditions correspond to $(1:0:0)$, $(0:1:0)$ and $(1/3:2/3:0)$, respectively. In terms of $\Psi^{\text{ini}}_{\alpha}$, this corresponds to
	
	\begin{centering}
		\begin{align}
		\Psi^{\text{ini}}_{\text{ND}}=\begin{pmatrix}
		1\\0\\0
		\end{pmatrix}, \quad
		\Psi^{\text{ini}}_{\text{MD}}=\begin{pmatrix}
		0\\1\\0
		\end{pmatrix},\quad
		\Psi^{\text{ini}}_{\text{PD}}&=\begin{pmatrix}
		\frac{1}{\sqrt{5}}\\\frac{2}{\sqrt{5}}\\0
		\end{pmatrix},
		\label{Eq:In_cond}
		\end{align}
	\end{centering}
	normalized such that the sum of probabilities is 1. 
	
	Our purpose in this section is to see observational differences between $H_0^{\text{EU}}$ and $H_0^{\text{LU}}$ in neutrino oscillations. In addition to that, we distinguish between inverted hierarchy (IH) for neutrinos masses, as well as the normal one (NH). This will result in a total of four cases for each initial neutrino composition eq.~\eqref{Eq:In_cond}: NH-EU, NH-LU, IH-EU and IH-LU. For instance, NH-EU corresponds to having neutrinos in the NH with today's rate of acceleration given by $H_0^{\text{EU}}$. In table~\ref{Table}, we list the values of the different neutrino parameters for both hierarchies as reported in~\cite{Review-Particle-Physics}. 
	
	\begin{center}
		\begin{table}
			\begin{center}
				\begin{tabular}{|c|c|c|}
					
					\hline
					Parameter&Normal Hierarchy (NH) &Inverted Hierarchy (IH)\\
					
					\hline
					$\sin^2(\theta_{12})$   & 0.307$\pm0.013$&0.307$\pm0.013$\\
					\hline
					$\Delta m_{21}^2$&  $(7.53\pm0.18)\times10^{-5}$eV${}^2$  &$(7.53\pm0.18)\times10^{-5}$eV${}^2$ \\
					\hline
					$\sin^2(\theta_{13})$ & $(2.18\pm0.07)\times10^{-2}$ & $(2.18\pm0.07)\times10^{-2}$\\
					\hline
					$\sin^2(\theta_{23})$& 0.545$\pm0.021$ & 0.547$\pm0.021$\\
					\hline
					$\Delta m_{32}^2$& $2.453\times10^{-3}$eV${}^2$  & $-2.546\times10^{-3}$eV${}^2$  \\
					\hline
					$\delta$& $1.36\pm0.36\pi$ rad(2$\sigma$)  & $1.36\pm0.36\pi$ rad(2$\sigma$)\\
					\hline
				\end{tabular}
			\end{center}
			\caption{Neutrino oscillation parameters used in the analysis as reported in~\cite{Review-Particle-Physics}}
			\label{Table}	
		\end{table}
	\end{center}
	
\begin{figure}[!tbp]
		\centering
		\includegraphics[width=\textwidth]{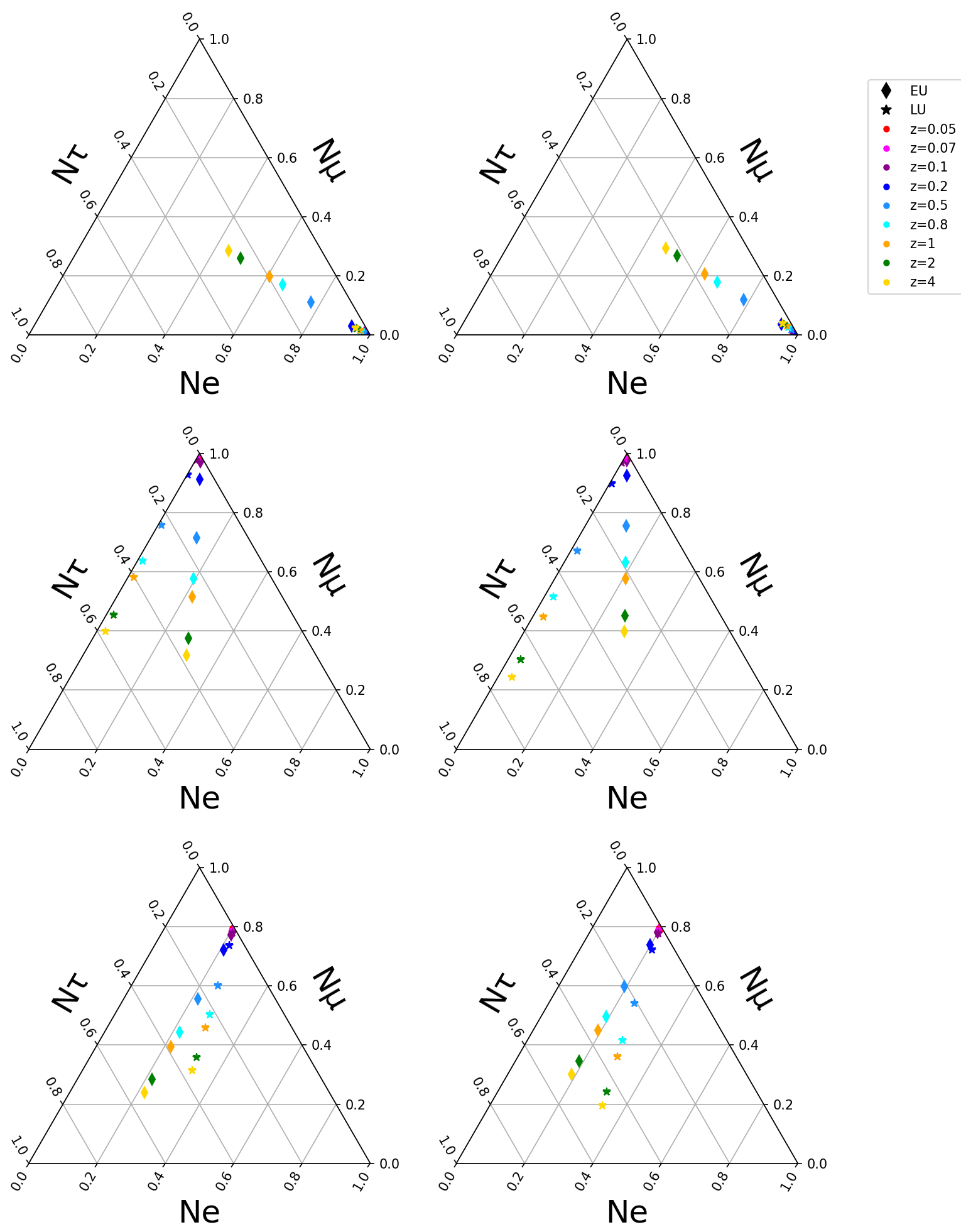}
		\caption{Ternary plots of ND (top), MD (middle) and PD (bottom) initial neutrino flavor decomposition, eq.\eqref{Eq:In_cond}, for NH (left) and IH (right). Diamond shaped points correspond to having $H_0^{\text{EU}}$ as today's rate of expansion, while star shaped ones for $H_0^{\text{LU}}$. Different colors correspond to emission redshifts, as given by the legend above. The python script to produce these is available in~\cite{GitHub}, which was written using~\cite{Ternary_Package}.}
		\label{fig:Ternary}
	\end{figure}

	One thing we can look at observationally is flavor ternary plots. These are triangular diagrams, with each side indicating the percentage of neutrinos from a certain flavor detected. In other words, each side corresponds to the probability of detecting neutrinos with certain flavor, with the sum being always equal to 1. These are shown in figure~\ref{fig:Ternary}. The first, second and third rows correspond to ND, MD and PD initial conditions, respectively. Moreover, the left side plots correspond to NH, while the right side ones to IH. Finally, in each diagram, different colors represent the indicated redshifts of emission, diamonds correspond to using $H_0^{\text{EU}}$ in the analysis of the previous section, while stars correspond to using $H_0^{\text{LU}}$.
	
	The first thing to note when looking at these diagrams is the difference between the left and right side ones for each IC. There is a slight distinction between hierarchies throughout their evolution with redshift. Therefore there is no degeneracy between NH and IH as neutrinos travel in an expanding universe, as expected. Second, once can notice an appreciable difference in evolution between different ICs, and therefore we see no degeneracy between them as well. Third, in every diagram, the distinction between EU and LU starts to become appreciable at around $z_e\sim 0.2$\footnote{On the other hand, there is a clear distinction between the two for ND initial conditions. The fact that there is very little change for ND in the LU case is a distinctive feature compared to the other cases.}. To give a concrete example on this distinction, let's look at the middle-right diagram of figure~\ref{fig:Ternary}, particularly the points corresponding to z=2. If at some point we detect at neutrino observatories, such as IceCube~\cite{IceCube}, neutrinos coming from a source with known redshift z=2, then we should find the flavor fraction given by the green star if the true value of $H_0$ is $H_0^{\text{LU}}$. However, if the detected flavor-fraction is given by the green diamond, and we are certain about the source's redshift, then we deduce that $H_0=H_0^{\text{EU}}$.
	%To avoid confusion, we will denote the transition probability when using $H_0^{\text{EU}}$ as $P^{\text{EU}}$, and a corresponding one for $H_0^{\text{LU}}$, $P^{\text{LU}}$.

\begin{figure}[tbp]
		\hspace{-4.2cm}
		\subfloat{\includegraphics[width=0.8\textwidth, keepaspectratio]{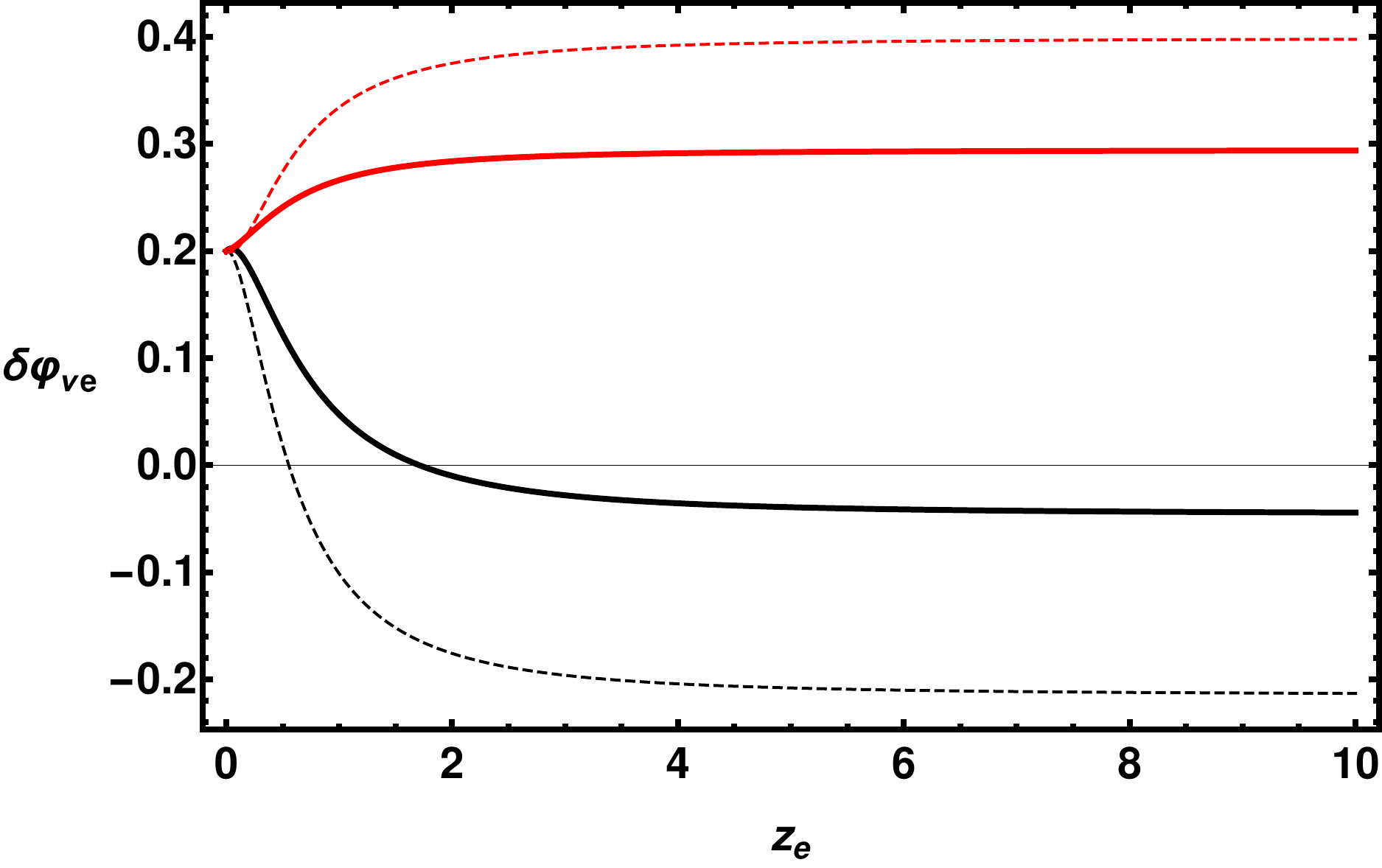}}\quad
		\subfloat{\includegraphics[width=0.8\textwidth, keepaspectratio]{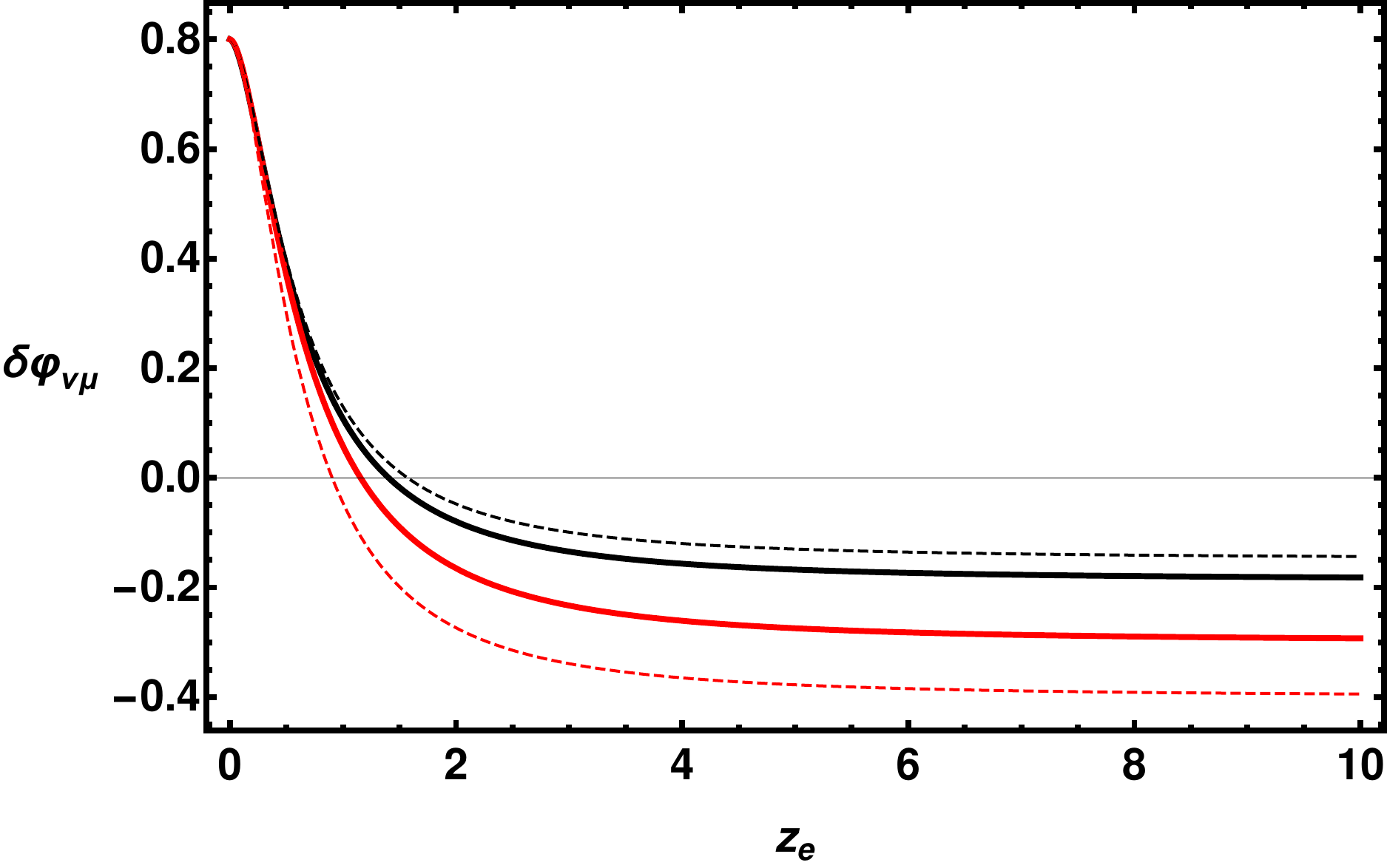}}
		
		\subfloat{\includegraphics[width=0.8\textwidth, keepaspectratio]{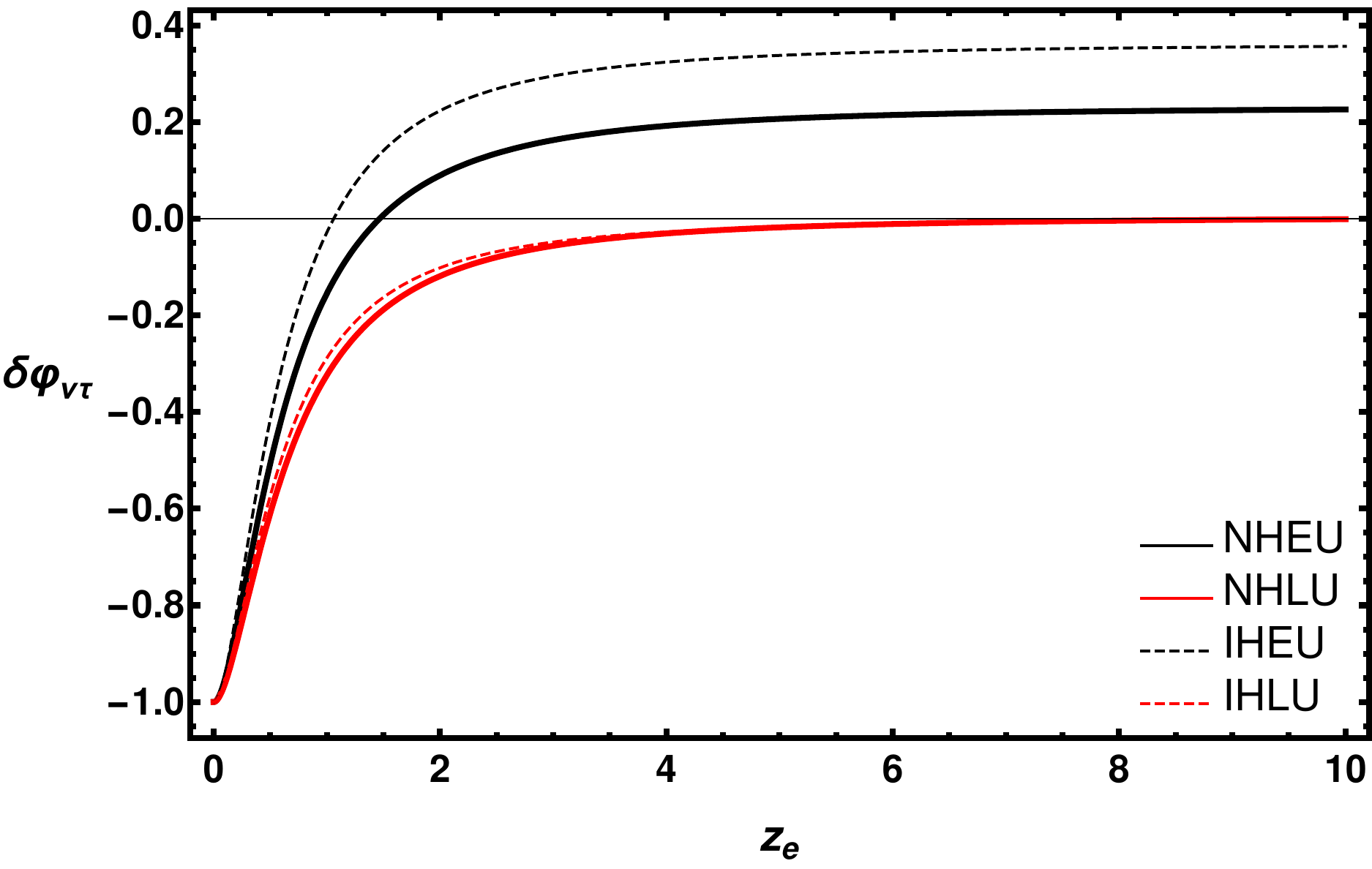}}
		\caption{Fractional difference, eq.~\eqref{Eq:Flux_Frac_Diff}, between observed and emitted neutrino fluxes as a function of redshift. The presented $\delta\varphi$s are for $\nu_e$(top-left), $\nu_{\mu}$(top-right) and $\nu_{\tau}$(bottom). Black curves correspond to having $H_0^{\text{EU}}$, while red ones to $H_0^{\text{LU}}$. On the other hand, solid lines refer to NH, while dashed ones to IH. The \texttt{Mathematica} script used to produce these is available in~\cite{GitHub}.}
		\label{fig:Flux_tot}
		
	\end{figure}

	Another observable that we can consider is the total neutrino flux. For a given flavor $\beta$, the total flux received at the detector, $\varphi_{\beta_0}$, is given by:
	\begin{align}
	\varphi_{\beta_0}&=\sum_{\alpha}^{ }P_{\alpha\beta}\varphi_{\alpha_e}\nonumber\\ &=\varphi_{\beta_e} +
	\sum_{\alpha; i<j}^{}\bigg[a_{\alpha\beta;ij}\sin^2\bigg(\frac{\Delta m_{ij}^2\Delta\lambda}{4}\bigg)+b_{\alpha\beta;ij}\sin\bigg(\frac{\Delta m_{ij}^2\Delta\lambda}{2}\bigg)\bigg]\varphi_{\alpha_e}
	\label{Flux}
	\end{align}
	where eq.~\eqref{Probability} has been used in the second line and $\varphi_{\alpha_e}$ is the flux at emission. When analyzing astrophysical neutrino fluxes, it is usually assumed that it takes an empirical form $\varphi_{\alpha_e}\sim AE_{\nu}^{-\gamma}$, where $A$ is a normalization constant and $\gamma$ is the spectral index, for any flavor~\cite{IceCube:Flux,IceCube:Flux2,Arguelles:Flux}. Therefore, the $\varphi_{\alpha_e}$s on the right hand side (r.h.s) of eq.~\eqref{Flux} can be factorized, allowing us to form a fractional difference, 
	\begin{equation}
	\delta\varphi_{\nu\beta}=\frac{\varphi_{\beta_0}-\varphi_{\beta_e}}{\varphi_{\beta_e}},
	\label{Eq:Flux_Frac_Diff}
	\end{equation} between the observed and emitted fluxes, figure~\ref{fig:Flux_tot}. Note that in these plots, the total flux for each flavor is being presented, i.e. summing over all ICs eq.~\eqref{Eq:In_cond}.

\begin{figure}[!tbp]
		\hspace{-4.2cm}
		\subfloat{\includegraphics[width=0.8\textwidth, keepaspectratio]{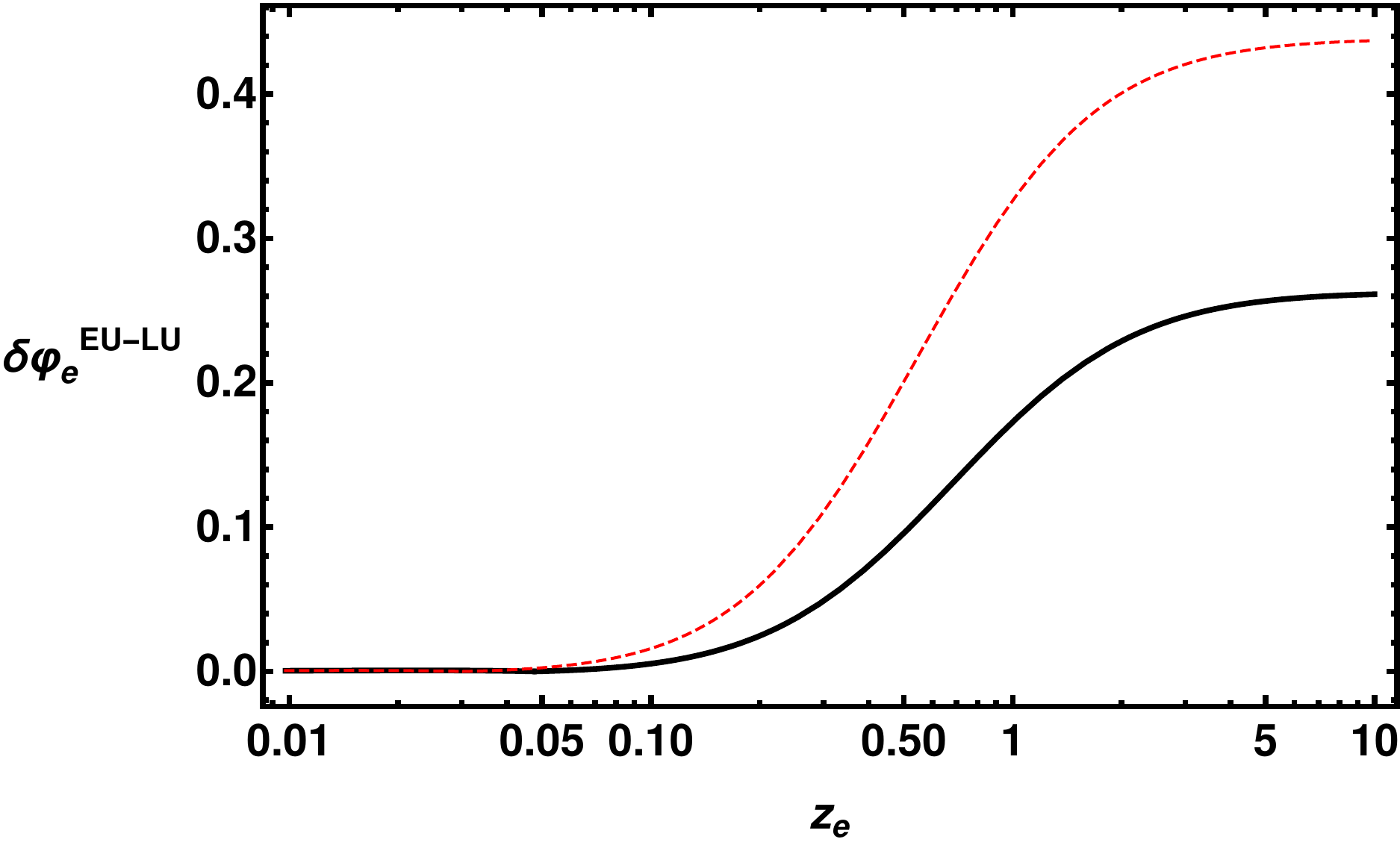}}\quad
		\subfloat{\includegraphics[width=0.8\textwidth, keepaspectratio]{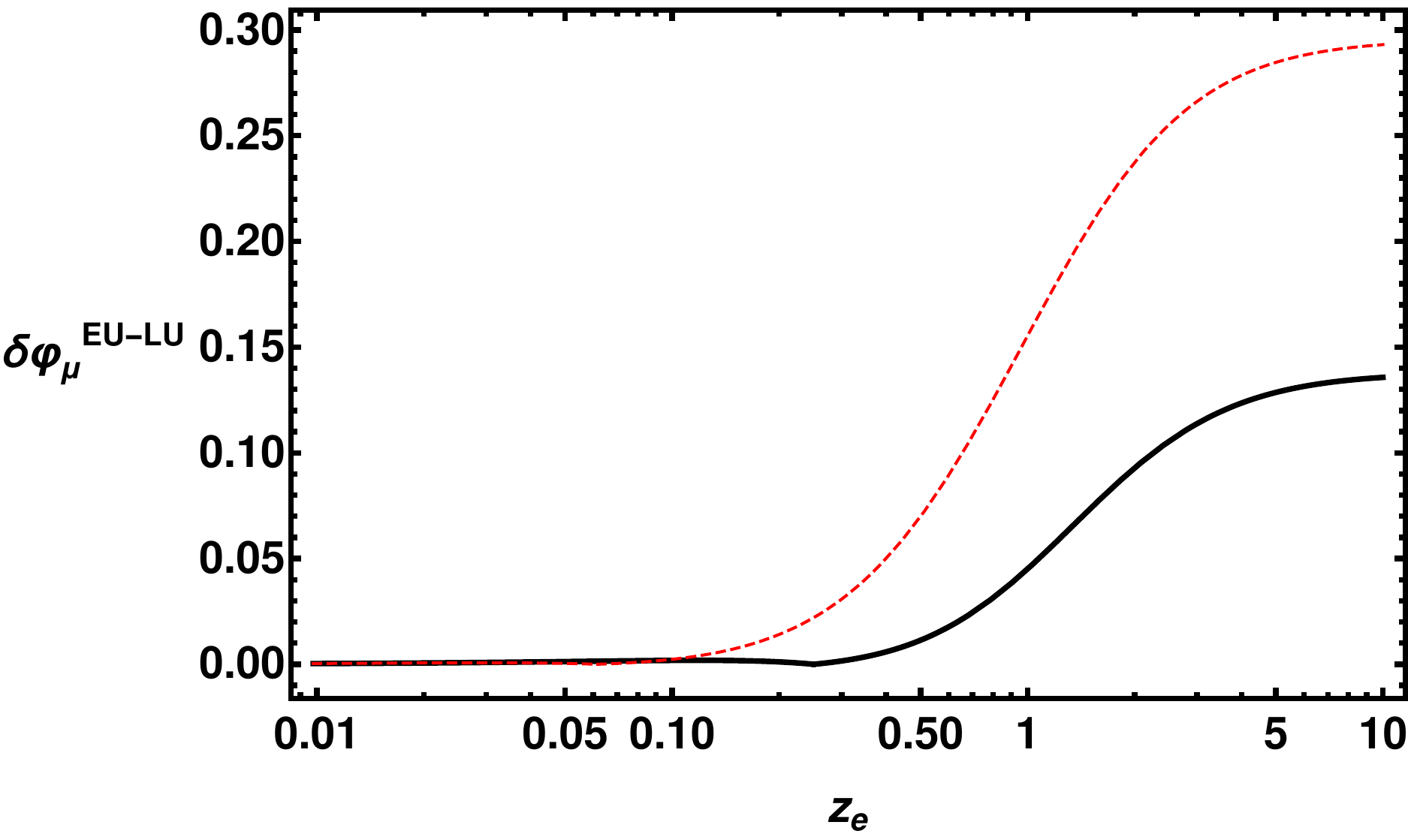}}
		
		\subfloat{\includegraphics[width=0.8\textwidth, keepaspectratio]{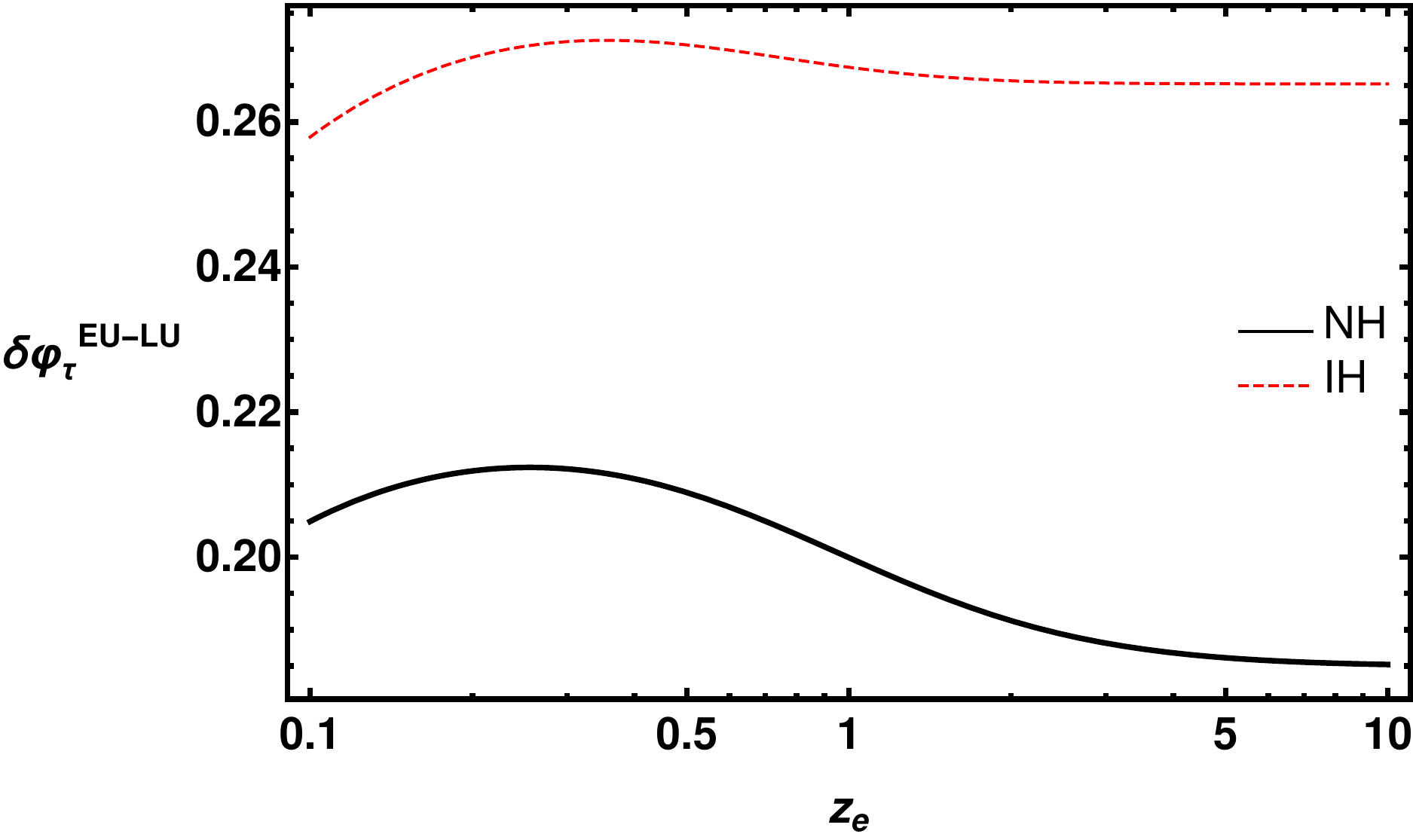}}
		\caption{Log-Linear plots of the fractional difference between EU and LU for each flavor flux as given in eq.~\eqref{Eq:Flux_EU-LU_Diff}. The differences presented in each plot are for both hierarchies and correspond to $\nu_e$(top-left), $\nu_{\mu}$(top-right) and $\nu_{\tau}$(bottom). The black-solid line correspond to having NH, while the red-dashed one to IH. The \texttt{Mathematica} script used to produce these is available in~\cite{GitHub}.}
		\label{fig:Flux_Diff_tot}
	\end{figure}

	Let us now make a few comments about these plots. First, all diagrams of figure~\ref{fig:Flux_tot} show noticeable differences between hierarchies and EU/ LU values of $H_0$. To see this more clearly, we plot in figure~\ref{fig:Flux_Diff_tot} the fractional difference between EU and LU for each of the quantities appearing in figure~\ref{fig:Flux_tot} and for both hierarchies. That is, we look at 
	\begin{equation}
	\delta\varphi_{\alpha}^{\text{EU}-\text{LU}}=\frac{\varphi_{\alpha_0}^{\text{EU}}-\varphi_{\alpha_0}^{\text{LU}}}{\text{Max}[\varphi_{\alpha_0}^{\text{EU}},\varphi_{\alpha_0}^{\text{LU}}]}
	\label{Eq:Flux_EU-LU_Diff}
	\end{equation}
	for each hierarchy and flavor $\alpha$, as a function of redshift. Even at relatively small redshifts ($z_e\sim 0.1$), using $H_0^{\text{EU}}$ or $H_0^{\text{LU}}$ makes a difference of a few \% on the flux received.
	
	Second, the starting values of figure~\ref{fig:Flux_tot}'s diagrams is related to the fact that our ICs eq.~\eqref{Eq:In_cond} are mainly of $\nu_e$ and $\nu_{\mu}$ type. As they evolve with redshift, neutrinos start changing flavor to one another. In particular, $\nu_e$ and $\nu_{\mu}$ are mostly transitioning to $\nu_{\tau}$, explaining the negative values of the top diagrams in figure~\ref{fig:Flux_tot}. Moreover, there is a $\nu_{\mu}\rightarrow\nu_{\tau}$ transition as well, which can be seen from the decreasing (increasing) character of the middle (bottom) diagram in the aforementioned figure. However, since we started with more $\nu_{\mu}$ than $\nu_e$, in addition to $\nu_{e,\mu}\rightarrow\nu_{\mu}$ being more dominant than the other transitions, then we have $\varphi_{\mu_0}>\varphi_{\mu_e}$.
	
	To give a more complete picture of how the difference between $H_0^{\text{EU}}$ and $H_0^{\text{LU}}$ affects neutrino oscillations, we can also look at the evolution of individual transition probabilities with redshift, as was done in~\cite{Khalifeh:2021ree}. However, in order to avoid clustering of diagrams, we report those in a GitHub repository~\cite{GitHub}.
	
	%We end this section by presenting the ternary plots for the case of ND ICs, for both hierarchies NH and IH in figure~\ref{fig:Ternary}. These plots contain the same information as the ones presented in figure~\ref{fig:Probability_ND}, but in a compact way. Each point represents the fraction of $\nu_{e},\ \nu_{\mu}$ and $\nu_{\tau}$ detected if the original composition is given by ND coming from the redshift corresponding to that point.
	
	\section{Conclusion}
	\label{Sec:Conc}
	
	With the persistence of a tension between early and late universe probes of today's expansion rate, $H_0$, using additional probes could shed some new light on the matter. In this work, which is an extension of previous ones~\cite{Khalifeh:2020bdg,Khalifeh:2021ree}, we demonstrated how neutrino oscillations can be such a probe.
	
	We considered a system of three-flavors neutrinos as spinors in a flat FRW universe, with a cosmological constant DE, $\Lambda$. We use neutrino parameters, table~\ref{Table}, and initial conditions eq.~\eqref{Eq:In_cond} to study the evolution of transition probabilities and neutrino fluxes with redshift. In particular, for each IC, figure~\ref{fig:Ternary} shows the detected flavor composition, distinguishing between $H_0^{\text{EU}}$ and $H_0^{\text{LU}}$ on the one hand, and between hierarchies on the other. Moreover, this distinction is presented for several redshifts of emission, demonstrating how the probability evolves with it. We can conclude from this that using $H_0^{\text{EU}}$ or $H_0^{\text{LU}}$ creates a difference of about 10\% on neutrino oscillations, starting from a redshift of emission of about 0.2. 
	
	Concerning detected neutrino fluxes, we consider the sum of all initial conditions in eq.~\eqref{Eq:In_cond} for each neutrino flavor $\nu_{e}, \nu_{\mu}$ and $\nu_{\tau}$. Fractional difference between detected and emitted fluxes is shown in figure~\ref{fig:Flux_tot}, for the four different combinations of hierarchies and early/late universe values of $H_0$. On the other hand, the fractional difference of the latter's effect on the fluxes is presented in figure~\ref{fig:Flux_Diff_tot}. The same conclusion previously reached applies here as well, showing the potential of using neutrino oscillations as a new probe of the Hubble tension.
	
	It is worth emphasizing that the considerations of this work are a mere generalization of neutrino oscillation studies to curved spacetime. No new entity or force have been added, rather a simple combination of distinct, well established, phenomena: neutrino oscillations and the Universe's accelerated expansion. Therefore, such an effect must be observed at some point in neutrino observatories~\cite{IceCube,IceCube-Gen2}, if it wasn't already in disguise. If this effect is not detected, even with the increasing performance of neutrino observatories~\cite{Boser:2013oaa}, then this could hint to new Physics in the neutrino or gravitational sectors.
	
	\section{Acknowledgment}
	We would like to thank Samuel Brieden for insightful discussions. The work of ARK and RJ is supported by MINECO grant PGC2018-098866-B-I00 FEDER, UE. ARK and RJ acknowledge ``Center of Excellence Maria de Maeztu 2020-2023" award to the ICCUB (CEX2019- 000918-M).

	\bibliography{biblio}
	\bibliographystyle{elsarticle-num}
\end{document}